\documentclass[nofootinbib,prd,eqsecnum,showpacs,showkeys,preprintnumbers,altaffilletter]{revtex4-1} 
\usepackage{amsmath}
\usepackage{amssymb}
\usepackage{graphicx}
\usepackage{amsfonts}
\usepackage{color}
\usepackage{bm}
\usepackage{mathrsfs}
\usepackage{url}

\usepackage[normalem]{ulem}
\usepackage{bigints}

\makeatletter

\newcommand*{\rom}[1]{\expandafter\@slowromancap\romannumeral #1@}

\newcommand{\be}{\begin{equation}}
  \newcommand{\ee}{\end{equation}}
\newcommand{\ben}{\begin{eqnarray*}}
  \newcommand{\een}{\end{eqnarray*}}
\newcommand{\bea}{\begin{eqnarray}}
  \newcommand{\eea}{\end{eqnarray}}
\newcommand{\bdm}{\begin{displaymath}}
  \newcommand{\edm}{\end{displaymath}}
\newcommand{\ba}{\begin{align}}
  \newcommand{\ea}{\end{align}}

\makeatother

\begin{document}

\title{Classical and quantum cosmology of the little rip abrupt event}

\author{Imanol Albarran $^{1,2}$}
\email{imanol@ubi.pt}

\author{Mariam Bouhmadi-L\'{o}pez $^{1,2,3,4}$}
\email{{\mbox{mbl@ubi.pt (On leave of absence from UPV and IKERBASQUE.)}}}

\author{{Claus Kiefer $^{5}$}}
\email{kiefer@thp.uni-koeln.de}

\author{Jo\~{a}o Marto $^{1,2}$}
\email{jmarto@ubi.pt}

\author{{ Paulo Vargas Moniz $^{1,2}$}}
\email{pmoniz@ubi.pt}

\date{\today}

\affiliation{
${}^1$Departamento de F\'{\i}sica, Universidade da Beira Interior, 6200 Covilh\~a, Portugal\\
${}^2$Centro de Matem\'atica e Aplica\c{c}\~oes da Universidade da Beira Interior (CMA-UBI), 6200 Covilh\~a, Portugal\\
${}^3$Department of Theoretical Physics University of the Basque Country UPV/EHU. P.O. Box 644, 48080 Bilbao, Spain\\
${}^4$IKERBASQUE, Basque Foundation for Science, 48011, Bilbao, Spain\\
${}^5${Institut f\"{u}r Theoretische Physik, Universit\"{a}t zu
  K\"{o}ln,  Z\"{u}lpicher Strasse 77, 50937 K\"{o}ln, Germany\\} 
}

\begin{abstract}

We analyze from a classical and quantum point of view
the behavior of the universe 
close to a little rip, which can be interpreted as a big rip sent
towards the infinite future. Like a big rip singularity, a little rip
implies the destruction of all bounded structure in the Universe
and is thus an event where quantum effects could be
important. We present here a new phantom scalar
field model for the little rip. 
The quantum analysis is performed in quantum
geometrodynamics, with 
the Wheeler-–DeWitt equation as its central equation. We find that the
little rip can be avoided in the sense of the DeWitt criterion, that
is, by having a vanishing wave function at the place of
the little rip. Therefore our analysis completes the answer to the
question: can quantum cosmology smoothen or avoid the divergent behavior  
genuinely caused by phantom matter?  
We show that this can indeed happen for the little rip, similar to the avoidance of a
big rip and a little sibling of the big rip.

\end{abstract}



\maketitle


\section{Introduction}

One of the most challenging problems in theoretical physics is the
formulation of a consistent quantum theory of gravity 
\cite{KieferQG,PMonizQC}. Such a theory is needed not only for conceptual
reasons, but also for understanding the origin of the Universe and the
structure of black holes. In our paper, we shall deal with quantum
cosmology, that is, the application of quantum theory to the Universe
as a whole. For this purpose, we shall use the conservative framework
called quantum geometrodynamics, with the Wheeler-DeWitt equation as
its central equation. This framework is
straightforwardly obtained by constructing 
quantum wave equations from which the Einstein equations can be
recovered in the semiclassical (WKB) limit \cite{GRG}. 

Besides these fundamental issues, we also encounter the problem to
explaining the observed acceleration of the
Universe. Phenomenologically, this is done by adding an ingredient
called dark energy (DE) \cite{AmendolaTsujikawa}. Some of the models
describing DE predict the occurrence of singularities beyond
big bang (or big crunch), occurring for example in the finite
future. Aside from DE singularities, there are also DE abrupt events like the little rip
\cite{Nojiri:2005sx,Nojiri:2005sr,Ruzmaikina,Stefancic:2004kb,BouhmadiLopez:2005gk,Frampton:2011sp,Bouhmadi-Lopez:2013nma,Brevik:2011mm}. We name them abrupt events rather than singularities because they occur at an infinite future cosmic time. Some of these models are in
accordance with current data \cite{Jimenez:2016sgs}. Since the presence of singularities
and abrupt events in a 
theoretical framework is an indication of its breakdown, we expect
quantum effects to be important there, too. A central
question is then whether those future singularities and abrupt events
can be avoided in quantum cosmology or not \cite{kolymbari}. This question will also be
addressed (and answered) for the models discussed in our paper. Naively, we would expect that at cosmological scales quantum effects
are important only in the early Universe, that is, on time scales of the
order of the Planck time, $t_{\rm P}$, and for distances 
related to the Planck length $l_{\rm P}$. This
naive belief is based on the fact that quantum theory is usually
important for small systems such as atoms or
molecules. Assuming the universality of the superposition
principle, quantum effects can occur at any scale, whenever
decoherence is negligible. This can happen even for the Universe as
a whole, for example, in the case of a classically recollapsing
Universe \cite{KZ95}, or in cases where singularities or abrupt
events are present in the classical theory, as is the case here.

Before proceeding further, we should clarify that among all DE
singularities and abrupt events, only three of them are intrinsic to
phantom DE, that is, within a relativistic model they happen if and only
if suitable phantom matter is present. These are the big rip, the
little rip, and the little sibling of the big rip. Consequently, if we
want to address the question: \textit{can quantum cosmology smoothen or
  avoid divergent behaviors caused by phantom matter}, we need to
quantize models that induce in the classical picture a big rip, a
little rip, or a little sibling of the big rip. These questions have
been partially addressed  in the quantum theory of
cosmological models 
with a big rip \cite{Dabrowski:2006dd,Albarran:2015tga} or a little
sibling of the big rip \cite{Albarran:2015cda}.
In this paper, we will complete the answer to these
questions by quantizing a classical model for the little rip. 
In addition and for completeness, we now recall  the
definition of 
the big rip, the little rip, and the little sibling of the big rip: 
	\begin{itemize} 
		\item Big rip singularity: It takes place  at a finite
                  cosmic time with an infinite scale factor where the
                  Hubble parameter and its cosmic time derivative
                  diverge 
                  \cite{Starobinsky:1999yw,Caldwell:2003vq,Caldwell:1999ew,Carroll:2003st,Chimento:2003qy,Dabrowski:2003jm,GonzalezDiaz:2003rf,GonzalezDiaz:2004vq}.   
		\item Little rip: This case corresponds to an abrupt
                  event rather than a future space-time
                  singularity. The radius of the Universe, the Hubble
                  parameter and its cosmic time derivative all diverge
                  at an infinite cosmic time
                  \cite{Nojiri:2005sx,Nojiri:2005sr,Ruzmaikina,Stefancic:2004kb,BouhmadiLopez:2005gk,Frampton:2011sp,Bouhmadi-Lopez:2013nma,Brevik:2011mm}. In 
                  addition, all the structure in the Universe would be 
                  ripped apart in a finite cosmic time
                  \cite{Frampton:2011sp}. This kind of behavior was
                  first found in \cite{Ruzmaikina} within the context
                  of a four-dimensional modified gravity model and later
                  on in \cite{BouhmadiLopez:2005gk} in an induced
                  gravity brane-world model. In these two papers
                  \cite{Ruzmaikina,BouhmadiLopez:2005gk} the little
                  rip was induced by pure geometrical effects. The name
                  little rip was coined in \cite{Frampton:2011sp},
                  where the abrupt event was induced by matter with
                  the equation of state (\ref{pressure}). This
                  equation of state was analyzed previously, as far as
                  we know, in \cite{Stefancic:2004kb} (see also
                  \cite{Nojiri:2005sx,Nojiri:2005sr}). 
		\item Little sibling of the big rip: This case again
                  corresponds to an abrupt event rather than a future
                  space-time singularity. At this event, the Hubble
                  rate and the scale factor blow up but the cosmic
                  derivative of the Hubble rate does not
                  \cite{Bouhmadi-Lopez:2014cca}. Therefore, this
                  abrupt event takes place at an infinite cosmic time
                  where the scalar curvature 
                    diverges. In addition, 
                  even though the event seems to be harmless as it
                  takes place in the infinite future, the bound
                  structures in the universe would be unavoidably
                  destroyed in a finite cosmic time from now. This was
                  first analyzed in \cite{Bouhmadi-Lopez:2014cca}
                  within a classical set-up. 		
	\end{itemize}

Our paper is organized as follows. In Sec. \ref{Sec2}, we review and
summarize some of the known results about the little rip abrupt event
from a classical point of view and as induced by a specific equation
of state (c.f. Eq.~(\ref{pressure})). In Sec. \ref{Sec3} and  for
later convenience, we introduce as well a scalar field suitable to
describe the nowadays late-time acceleration of the universe and
are simultaneously able to induce a little rip asymptotically in the
presence and absence of dark matter (DM). In Sec. \ref{WDWsection}, we
present and solve the Wheeler-DeWitt equation for the models given in
Secs. \ref{Sec2} and \ref{Sec3}. We show in both cases the
existence of solutions to the Wheeler-DeWitt equation that avoid the little
rip. Finally, in Sec. \ref{Sec5} we present our conclusions.
In addition, we include { Appendices \ref{appendix} and \ref{justapp},  in which we prove the validity
of the approximations used in Sec. \ref{WDWsection}, and Appendix \ref{appendix2}, where the Symanzik scaling behavior is presented as an alternative method to analyze the scalar field eigenstates.}    


\section{A brief review of the little rip event}
\label{Sec2}

In our paper, we shall employ a
Friedmann-Lema\^{\i}tre-Robertson-Walker (FLRW) model with flat
spatial sections (that is, choosing $k=0$). 
The little rip is obtained in this framework by introducing a perfect
fluid with equation of state
\cite{Stefancic:2004kb,Frampton:2011sp}\footnote{For previous work on this
kind of abrupt event, see
\cite{Nojiri:2005sx,Nojiri:2005sr,Ruzmaikina,Stefancic:2004kb,BouhmadiLopez:2005gk,Frampton:2011sp,Bouhmadi-Lopez:2013nma,Brevik:2011mm}.}  
\begin{equation}\label{pressure}
p_d=-\rho_d-A\sqrt{\rho_d},
\end{equation}
where $A$ is a positive constant, and $\rho_d$ and $p_d$ are the
energy density and the pressure of this fluid, respectively. The
subscript $d$ stands for DE, since the
observed acceleration of the Universe could be described by
(\ref{pressure}) \cite{Frampton:2011sp}. The constant $A$ has the
physical dimension of a square root of density and can thus be written
as
\be
\label{rhostar}
A=:\sqrt{\rho_*},
\ee
with a characteristic density $\rho_*$. 

Since we are interested in the description of the little rip, and
since this event occurs in the infinite future, we are
mainly interested in the asymptotic behavior of this model. This can
be addressed using standard cosmological equations. We first employ
the Friedmann-Lema\^{\i}tre equation
\begin{equation}\label{Friedman}
H^2=\frac{8\pi G}{3}\rho,
\end{equation}
where $H\equiv\dot{a}/a$. Here and in the following, we use units with
$c=1$. 
The energy density can be described as the sum of the contribution of
the different matter components in the Universe: 
\begin{equation}\label{eq:energdens}
{\rho}=\frac{3H_0^2}{8\pi G}\sum_j \Omega_j (a)\equiv \rho_{\rm
  c}\sum_j \Omega_j (a) ,
 \qquad  \qquad \Omega_{j0}=\frac{8\pi
   G}{3H_0^2}\rho_{j0}\equiv\frac{\rho_{j0}}{\rho_{\rm c}} ,
\end{equation}
where $\Omega_j$ denotes the energy density fraction of each component
of the Universe and $\rho_{\rm c}$ is the critical density;
the index $0$ means that the corresponding quantity is evaluated at
present time. At very late times, we can disregard
the contribution of DM to the total energy density; however, it
corresponds to a significant part of the 
present content. Fixing accurate values for the model
parameters $A$ and $\Omega_{d0}$ requires the imposition of
observational constraints. We will assume simply that
$0<\Omega_{d0}<1$, for the precise value has no affects on our 
analysis. Within this approximation, (\ref{Friedman}) contains just a
single term corresponding to the energy density of DE.  

The second equation that we use is the energy conservation equation,
from which one gets \cite{Frampton:2011sp} 
\begin{equation}\label{rho}
\rho_d=\rho_{d0}\left[\frac{3A}{2\sqrt{\rho_{d0}}}
\ln\left({\frac{a}{a_0}}\right)+1\right]^2, 
\end{equation}
where $a_0$ is an integration constant that we set equal to the current
size of the Universe. Therefore, after integrating (\ref{Friedman}),
the asymptotic behavior of the scale factor with respect to 
cosmic time reads \cite{Frampton:2011sp} 
\begin{equation}
\frac{a}{a_0}\sim\exp{\left[\beta\left(e^{\alpha t}-1\right)\right]},
\qquad \textrm{where} \qquad {\color{black}\alpha\equiv\sqrt{6\pi G}A}, \qquad
\beta\equiv\sqrt{\frac{\Omega_{d0}}{6\pi G}}\frac{H_0}{A}. 
\end{equation}
 The little rip event happens for large values of $a$, where $\rho$
 and $p$ blow up, and therefore also $H$ and $\dot{H}$. Notice that,
 unlike the big rip, the little rip event is reached in infinite
 cosmic time.  
 

\section{The little rip as induced by a scalar field}
\label{Sec3}

For later convenience, we map the perfect fluid with equation of state
(\ref{pressure}) to a scalar field, $\phi$. As the constant $A$ must
be positive for (\ref{pressure}) to induce a little
rip, the mapping to a scalar field entails a 
phantom character for the field. Consequently, we can write the
kinetic energy  and 
potential of the scalar field as 
\begin{equation}\label{phidot}
\dot{\phi}^2=-(\rho+p),
\end{equation}
\begin{equation}\label{potential}
V=\frac{1}{2}\left(\rho-p\right).
\end{equation}
Inserting (\ref{rho}) and (\ref{pressure}) in (\ref{phidot}), we get
\begin{equation}\label{phidotsquare}
\dot{\phi}^2=A\rho_d^{\frac{1}{2}}=\left\vert\frac{3A^2}{2}\ln\left({\frac{a}{a_0}}\right)+A\sqrt{\rho_{d0}}\right\vert.
\end{equation}
Introducing the new variable 
\begin{equation}\label{tau}
x \equiv\ln\left({\frac{a}{a_0}}\right) ,
\end{equation}
we can express $\dot{\phi}$ as
\begin{equation}\label{phidottau}
\dot{\phi}=\frac{d\phi}{d x} H.
\end{equation}
We now treat in separate subsections the cases without and with DM.

\subsection{Disregarding dark matter}

 Using (\ref{phidotsquare}) and (\ref{Friedman}), we can write
\begin{equation}\label{dphidtau}
{\color{black}{d\phi}=\frac{\dot{\phi}}{H}d x={\pm}\frac{\sqrt{3}}{\kappa}\left(\frac{\Omega_{*}}{\Omega_{d0}}\right)^{\frac{1}{4}}\frac{dx}{\left\vert\frac{3}{2}\sqrt{\frac{\Omega_{*}}{\Omega_{d0}}}x+1\right\vert ^{\frac{1}{2}}}}, 
\end{equation}
where $\kappa^2\equiv 8\pi G$
 and
$\Omega_{*}\equiv\left({A\kappa}/{\sqrt{3}H_0}\right)^2
\equiv \rho_*/\rho_{\rm c}$. The latter
denotes a critical energy density fraction which is related with  the
model parameter $A$ and quantifies the deviation of a DE model based
on (\ref{pressure}) from the standard $\Lambda$CDM model, that is, the
smaller is 
$\Omega_{*}$, the closer we are to the $\Lambda$CDM model. Notice that
the expression (\ref{dphidtau}) is only valid  asymptotically, for we
have disregarded the contribution of DM which will red-shift
quickly in the future and thus become negligible compared to DE.
Finally, from integrating (\ref{dphidtau}) we find {\color{black}(for $\Omega_{*}\neq0$)}
\begin{equation}\label{phi tau}
{\color{black}\phi(x)={\pm}\frac{4}{\sqrt{3}\kappa}\left(\frac{\Omega_{d0}}{\Omega_{*}}\right)^{\frac{1}{4}}\left\vert\frac{3}{2}\sqrt{\frac{\Omega_{*}}{\Omega_{d0}}}x +1\right\vert^\frac{1}{2}
\textrm{sign} \left(\frac{3}{2}\sqrt{\frac{\Omega_{*}}{\Omega_{d0}}}x +1\right).}
\end{equation}
{\color{black}We have chosen the integrations constants, $\phi_{*}$ and $x_{*}$
such that}
\begin{equation}\label{phi tau const}
{\color{black} \ \phi_{*}={\pm}\frac{4}{\sqrt{3}\kappa}\left(\frac{\Omega_{d0}}{\Omega_{*}}\right)^{\frac{1}{4}}\left\vert\frac{3}{2}\sqrt{\frac{\Omega_{*}}{\Omega_{d0}}}x_{*} +1\right\vert^\frac{1}{2}
\textrm{sign} \left(\frac{3}{2}\sqrt{\frac{\Omega_{*}}{\Omega_{d0}}}x_{*} +1\right).}
\end{equation}
In addition,  we have selected
$x_{*}$  to be large enough to ensure the validity of the
approximation made in (\ref{dphidtau}); that is, we are far
enough in the future such that the DM component can be
ignored in the Friedmann equation. For practical purpose, we
select $x_{*}=1.17$, where the matter energy density is two orders
of magnitude smaller than the DE density. Therefore, $x_{*}$
is large enough for the Universe to be in an almost total DE
domination phase. This numerical value is not crucial for this subsection, but it has to be fixed in the next subsection where numerical calculations are required and therefore a fixed value of $x_{*}$ is needed. In addition, our results do not change by imposing larger values of $x_{*}$.
Finally, the function
$\textrm{sign}\left(x\right)$ is the sign function, that is
\begin{equation}\label{sign}
\textrm{sign} \left(x\right) = \left\{\begin{array}{rcl}
 -1&\mbox{if}& x<0 \\
0&\mbox{if}& x=0 \\
1 & \mbox{if}& x>0
\end{array}\right..
\end{equation}

{\color{black}As mentioned before, the equation of state shown in Eq.(\ref{pressure}) describes a deviation from the  standard $\Lambda$CDM model through the parameter $A$. Therefore, for a vanishing  parameter $A$, the expected classical trajectory, $\phi\left(x\right)$, is characterized  by a constant, i.e. $d\phi=0$. This result  can be recovered   by taking the limit  $\Omega_{*}\rightarrow0$ in Eq.(\ref{dphidtau}), however, after the  integration done in Eq.(\ref{phi tau}) the  outcome  is not well defined for the limit $\Omega_{*}\rightarrow0$ (notice that $\phi_{*}$ could blows up in this case). To get a  suitable  expression for small values of  $\Omega_{*}$, we perform  a Taylor expansion up to first order of the general integral of Eq.(\ref{dphidtau}), which reads
\begin{equation}\label{phi tau small}
\phi(x)-\tilde{\phi}_*\simeq{\pm}\frac{\sqrt{3}}{\kappa}\left(\frac{\Omega_{*}}{\Omega_{d0}}\right)^{\frac{1}{4}}\left(x-\tilde{x}_*\right),
\end{equation}
where in this case, we have chosen $\tilde{\phi}_{*}$ and $\tilde{x}_*$ in such way that:
\begin{equation}\label{phi tau small const}
\tilde{\phi}_{*}={\pm}\frac{\sqrt{3}}{\kappa}\left(\frac{\Omega_{*}}{\Omega_{d0}}\right)^{\frac{1}{4}}\tilde{x}_*.
\end{equation}
This result will be used later to determine  the potential $V\left(\phi\right)$. }

In the little rip not only the scale factor gets very large, but also
the scalar field $\phi$, see Fig.~1. From now on we will focus on this
regime.

\begin{figure}[h!]
\centering\includegraphics[width=8cm]{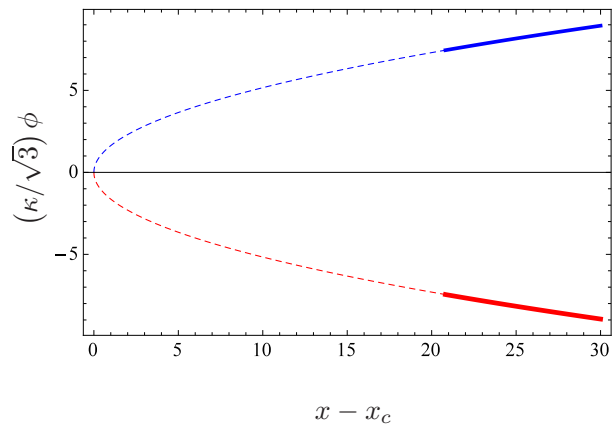}\caption{%
\small{Plot of the scalar field, $\phi$, versus $x\equiv\ln(a/a_0)$ {\color{black} where $x_c=-2\sqrt{\Omega_{d0}}/3\sqrt{\Omega_{*}}$. This plot is  valid 
for $\Omega_{*}\neq0$ since $x_c$ is not well defined for a vanishing 
$\Omega_{*}$, i.e.  a vanishing $A$. The solution (\ref{phi tau}) gives 
two branches, one above $\phi =0$ (blue color) and another below $\phi =0$ (red color)} The dashed curve describes a 
realm where the neglected DM contribution is important, while the solid 
lines describes a regime where  we assume a complete DE domination. 
We disregard the solutions for $x<x_c$ as our approximation breaks down there. 
Therefore, only the solid lines are physically relevant for our purpose.}%
}
\end{figure}

{ \color{black}For the case of $\Omega_{*}\neq0$, since the function  (\ref{phi tau}) is invertible, we can
consequently write $x=x(\phi)$,}
\begin{equation}\label{tauphi}
x=\frac{\kappa^2}{8} \phi^2-\frac{2}{3}\sqrt{\frac{\Omega_{d0}}{\Omega_{*}}},\qquad \textrm{for} \qquad 0< \frac{3}{2}\sqrt{\frac{\Omega_{*}}{\Omega_{d0}}}x +1,
\end{equation}
\begin{equation}\label{tauphi2}
x=-\frac{\kappa^2}{8} \phi^2-\frac{2}{3}\sqrt{\frac{\Omega_{d0}}{\Omega_{*}}}, \qquad \textrm{for} \qquad \frac{3}{2}\sqrt{\frac{\Omega_{*}}{\Omega_{d0}}}x +1<0.
\end{equation}
Once we have the relation between the potential and the energy density, (\ref{potential}), {\color{black}we can write  the potential in terms of $x$, that is
 \begin{equation}\label{potx}
V\left(x\right)=\rho_{d0}\left[\frac{3}{2}\sqrt{\frac{\Omega_{*}}{\Omega_{d0}}}x+1\right]^2+\frac{3H_0^2}{2\kappa^2}\sqrt{\Omega_{d0}\Omega_{*}}\left\vert\frac{3}{2}\sqrt{\frac{\Omega_{*}}{\Omega_{d0}}}x+1\right\vert.
\end{equation}
As can be seen, for a vanishing $\Omega_{*}$, the potential becomes constant as expected within the $\Lambda$CDM paradigm, i.e. $V=\rho_{d0}$. Using  Eq.(\ref{tauphi}) in the later expression, the potential shows a quadratic dependence on the scalar field:
}
\begin{equation}\label{potphi}
V\left(\phi\right)=b_1\phi^4+b_2\phi^2,
\end{equation}
where the constants $b_1$ and $b_2$ are defined as 
\begin{eqnarray}\label{b1b2}
b_1\equiv\frac{27}{256}\kappa^2H_0^2\Omega_{*}, \qquad
b_2\equiv\frac{9}{32}H_0^2\Omega_{*}. 
\end{eqnarray}
{\color{black}On the one hand, notice} that $b_1$ has physical dimension of an inverse mass times
length (and is thus dimensionless in natural units where $\hbar=1$ and $c=1$),
 while $b_2$ has dimension of an inverse length squared (mass squared
 in natural units). 
As was mentioned above, (\ref{phi tau}) does not take
into account the contribution of DM; therefore, the result
shown in Fig.~1 is only valid for very large values of the scale
factor.  

{\color{black} On the other hand, for a vanishing  parameter $A$ the potential  given in (\ref{potphi}) cannot  show the expected constant value. This is not surprising as (\ref{potphi}) was deduced using (\ref{phi tau}) which is not valid for $A=0$. To recover this solution it is necessary to replace  in Eq.(\ref{potx}) the expression obtained in Eq.(\ref{phi tau small}) for small values of $\Omega_{*}$. In fact, on that case, we obtain
\begin{equation}\label{Vasmall}
V\left(\phi\right)\simeq\rho_{d0}\left[\frac{\sqrt{3}\kappa}{2}\left(\frac{\Omega_{*}}{\Omega_{d0}}\right)^{\frac{1}{4}}\phi+1\right]^2+\frac{3H_0^2}{2\kappa^2}\sqrt{\Omega_{d0}\Omega_{*}}\left\vert\frac{\sqrt{3}\kappa}{2}\left(\frac{\Omega_{*}}{\Omega_{d0}}\right)^{\frac{1}{4}}\phi+1\right\vert.
\end{equation}
As can be seen from the previous expression when $A\rightarrow 0$, $V\left(\phi\right)$ approaches a constant; i.e. the model in this case behaves as $\Lambda$CDM.}

\subsection{Including dark matter}

Just for completeness and to get an accurate solution also for small
values of $x$ (but still large enough to be in a matter domination
epoch after the radiation dominated epoch), it is necessary to
incorporate the DM contribution to the energy density budget
of the Universe. Following the same approach we used before,
(\ref{phidottau}) can be written as 
{\color{black}
\begin{equation}\label{phidotwithmatter}
{d\phi}={\pm}\frac{\dot{\phi}}{H}d x=\left\{\frac{\left\vert p_d\left(x\right)+\rho_d\left(x\right)\right\vert}{H^2}\right\}^{\frac{1}{2}}dx.
\end{equation}}
 The contribution of DM is here included in the Hubble
 parameter. The equation for $\phi\left(x\right)$ is now given
 by 
\begin{equation}\label{numphi}
\phi(x)={\pm}\frac{\sqrt{3}}{\kappa}\bigintss_{x_{*}}^{x}
\left\{\frac{\sqrt{\Omega_{*}\Omega_{d0}}\left\vert
      \frac{3}{2}\sqrt{\frac{\Omega_{*}}{\Omega_{d0}}}x+1
    \right\vert}{\Omega_{m0}e^{-3x}+\Omega_{d0}\left(\frac{3}{2}\sqrt{\frac{\Omega_{*}}{\Omega_{d0}}}x+1\right)^2}\right\}^{\frac{1}{2}}dx
+\phi_{*}. 
\end{equation}
The integral in (\ref{numphi}) cannot be solved analytically;
therefore, we have performed a numerical integration in which the
integration constant $\phi_{*}$ was fixed as after (\ref{phi tau}) to
the value {\color{black} imposed in Eq.(\ref{phi tau const})}. In this way, we ensure that the
approximated model and the numerical solution are equal at the point
$x_{*}$ as long as $x_{*}$ is large enough. For practical purpose, we
select $x_{*}=1.17$, where the matter energy density is two orders
of magnitude smaller than the DE density. Therefore, $x_{*}$
is large enough for the Universe to be in an almost total DE
domination phase. Figure~2 shows $\phi(x)$.  

\begin{figure}[h!]
\includegraphics[width=8.3cm]{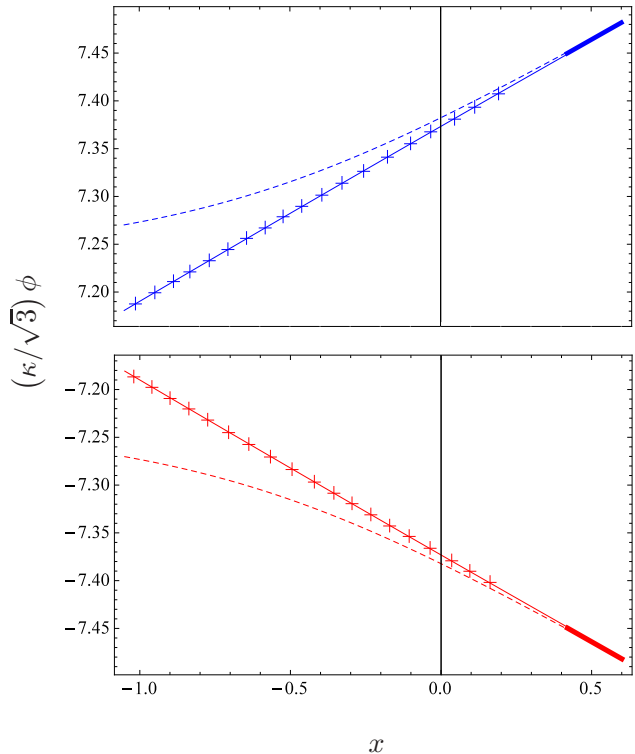}
\caption{\small{Plot of the rescaled scalar field,$\left(\kappa/\sqrt{3}\right)\phi$, versus $x$, 
the logarithmic scale factor. The solution (\ref{numphi}) has two branches which we have drawn 
as dashed lines {\color{black}in bottom (red) and upper (blue) panels}. These lines take into account
DM contribution. The solid blue and red lines correspond to the solution 
(\ref{phi tau}) where DM is neglected. All the plots have been obtained for $x_*=1.17$. 
For practical purpose, we see that for values larger than $x=0.42$ 
(i.e. the energy density of DM is $10$ times smaller than that of DE), 
the difference between the two solutions (inclusion of DM and exclusion of DM) 
is almost negligible. For values of $x$ smaller than $x=0.42$, the approximated solution starts to show a relevant deviation from the exact solution and we have drawn in this case the approximated solution as a curve with crosses. In addition, we have fixed the other constants as $H_0=70.1 \ \textrm{km}\ \textrm{s}^{-1}\textrm{Mpc}^{-1}$, $\Omega_{m0}=0.274$, and $A\kappa=3.46\cdot 10^{-3} \textrm{Gyr}^{-1}$ according to the best fit obtained in \cite{Frampton:2011sp}.}}

\label{fig2a}
\end{figure}

Once we have obtained the solution for the scalar field, we get the numerical
solution for the  potential $V\left(\phi\right)$, which  also takes
into account the DM contribution. We compare the obtained
potential with the approximated potential (which neglects DM)
in Fig.~3.

\begin{figure}[h!]
\includegraphics[width=8cm]{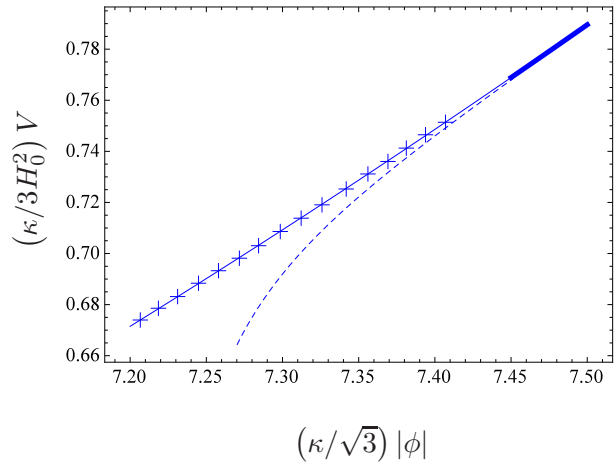}
\caption{\small{Plot of the dimensionless potential,$\left(\kappa/3H_0^2\right)V$, versus the absolute value of the scaled scalar field, $\left(\kappa/\sqrt{3}\right)\left\vert\phi\right\vert$. The dashed curve takes into account the presence of DM, while the solid line neglects it. In consistency with the other plots, we take $x_{*}=1.17$. The deviation becomes significant when $\left(\kappa/\sqrt{3}\right)\left\vert\phi\right\vert<7.45$ (drawn as thin curve with crosses), that is, $x<0.42$, corresponding to the energy density of DM being $10$ times smaller than that of DE.}}

\label{fig3}
\end{figure}
Because DM is completely negligible at late times, the
quantum analysis of the little rip is 
unaffected by it. We will thus neglect DM from now on. 
 


\section{Wheeler-DeWitt equation}\label{WDWsection}

The canonical formulation of general relativity leads to four local
constraints. If quantization is performed in the Dirac sense, they
turn into the Wheeler-DeWitt (WDW) equation and the quantum
diffeomorphism constraints \cite{KieferQG}. Making from the outset a
FRLW ansatz, as we do here, only the WDW equation in the form of one
partial differential equation remains. To be concrete, the classical
metric is (for a flat spatial metric) given by
\begin{equation}\label{metric}
g_{\mu\nu}dx^{\mu}dx^{\nu}=-N^2 (t) dt^2 +a^2(t) \left[dr^2+r^2
  d\theta^2 + r^2\sin^2(\theta) d\varphi^2\right], 
\end{equation}
where $N$ is the  lapse function, $a$ is the size of the Universe and
$t$ is the  cosmic time. Choosing, in addition, a perfect fluid 
and a set of minimally coupled scalar fields, the total
action consisting of Einstein-Hilbert action and matter action reduces to 
\cite{PMonizQC,KieferQG,Dabrowski:2006dd,Kamenshchik:2007zj,Albarran:2015tga,Albarran:2015cda}  
\begin{equation}\label{totalac}
\textrm{S}=\int\mathcal{L}\:\:dt, \qquad
\mathcal{L}=\frac{{\mathcal V}_0}{2\pi}\left[-\frac{3\pi}{4GN}\dot{a}^{2}a-2\pi^{2}Na^{3}
\left({\rho}-\frac{1}{2N^2}\sum_i  
  l_i{\dot{\phi}}^2+V\left(\phi_1,.. \phi_n\right)\right)\right], 
\end{equation}
where $l=1,(-1)$ for standard (phantom) scalar fields. The constant
${\mathcal V}_0$ stands for the volume of the three-dimensional
spatial sections for $a=1$. As we are dealing with spatially flat
sections, it is implicitly assumed  that we either choose a torus
compactification with the correct volume or we leave the volume open,  that is, choose a
reference volume ${\mathcal V}_0$. We take the second
option.  Please note that this leads only to a constant factor multiplying the 
Lagrangian and, therefore, will not affect the results presented below. 

Starting from the Lagrangian (\ref{totalac}),
the conjugate momenta can be calculated and the classical Hamiltonian
can be obtained from a Legendre transformation. The classical
constraint $\mathcal{H}=0$ then becomes after quantization
\cite{PMonizQC,KieferQG}  
\begin{equation}\label{H=0}
\hat{\mathcal{H}}\Psi=0.
\end{equation}
This equation enables an appropriate quantum approach where the wave
function of the Universe $\Psi$ depends on the degrees
of freedom used to describe the physical system under study,
that is, the configuration space.  We will
address quantization by means of: $(i)$ a
single degree of freedom corresponding to the scale factor, where the
matter content is given by a perfect fluid (with a known equation of
state corresponding to (\ref{pressure})), and $(ii)$ two degrees
of freedom corresponding to the scale factor and a scalar field (which
portrays the matter content).  

In the first approach, the scale factor is the only independent
variable. This is certainly a very simple model, but it is interesting
enough to study the behavior of the wave function near singularities. 
In the absence of a full quantum gravity framework it is, of course,
an open question what the correct criterion of singularity avoidance
is. A useful heuristic criterion is the one introduced by DeWitt in
1967 \cite{DeWitt}; it states that the wave function should vanish at
the place of the classical singularity. This criterion was
successfully applied to a variety of cosmological models, see 
\cite{PMonizQC,KieferQG,kolymbari} and
\cite{BouhmadiLopez:2009pu,Bouhmadi-Lopez:2013tua,Albarran:2015tga,Albarran:2015cda,KKK16}.     

In the second approach, an approximation describing the matter content
by a scalar field yields a suitable framework with an additional
degree of freedom.  We move from the classical trajectory,
$\phi=\phi\left(a\right)$,  to the corresponding quantum analog
where the wave function is defined over the configuration space
($a,\phi$). In this way, the quantum nature arises and gains
significance close to the singularity, once the quantum effects become
important. Here, again, the DeWitt criterion is useful as a heuristic
device.


\subsection{Wheeler-DeWitt equation with a perfect fluid}
In this subsection, we will implement the quantization in the simplest
way, which consists in describing the  matter content as a perfect
fluid with a given equation of state. Therefore, the energy density
can be written in terms of the scale factor, which is the single
variable within this analysis. The Lagrangian for this model reads
\cite{KieferQG,PMonizQC,BouhmadiLopez:2004mp,Albarran:2015tga} 
\begin{equation}
\mathcal{L}=-\frac{3\pi}{4GN}\dot{a}^{2}a-2N\pi^{2}a^{3}{\rho}\left(a\right).
\end{equation}
The conjugate momentum is
\begin{equation}\label{canonicalmomentum}
\pi_{a}\equiv\frac{\partial \mathcal{L}}{\partial\dot{a}}=-\frac{3\pi }{2G}N a\dot{a},
\end{equation}
and the Hamiltonian reads
\begin{equation}\label{eq:wkbHamiltconstraint}
\mathcal{H}=-N\frac{G}{3\pi }\frac{\pi_{a}^{2}}{a}+2\pi^{2}Na^3\rho\left(a\right).
\end{equation}
For the sake of simplicity, we introduce the new constants 
\begin{equation}\label{eta}
\eta\equiv\frac{\pi {a}_0^3H_0}{G\hbar}, \qquad b\equiv
\frac32\sqrt{\frac{\Omega_{*}}{6\Omega_{d0}}}. 
\end{equation}
The exact form of the WDW equation depends on the chosen factor
ordering. We shall employ two different such orderings in order to
study its influence.
\subsubsection{First quantization procedure:
  $a\mathcal{\hat{H}}(a,\hat{\pi}_{a})\psi(a)=0$} 
We choose here \cite{Dabrowski:2006dd}
\begin{equation}\label{eq:momentum}
\hat{\pi}_{a}^2=-\hbar^2\partial_a^2.
\end{equation}
Employing this in the quantum version of (\ref{eq:wkbHamiltconstraint})
 and multiplying the result by $\left[3 \pi a/G \hbar^2N\right]$, we get
\begin{equation}\label{Hamilq1}
\frac{3 \pi}{G \hbar^2}\frac{a}{N}\mathcal{\hat{H}}=\partial_a^2+\frac{6\pi^3 }{\hbar^2 G}a^{4}\rho\left(\frac{a}{a_0}\right).
\end{equation}
In order to get a dimensionless WDW equation, we will rescale the
scale factor and its partial derivative as  
\begin{equation}\label{scalefactor}
u\equiv\frac{a}{a_0}, \qquad \partial_u^2\equiv{a_0^2}\partial_a^2.
\end{equation}
After carrying the change of variable introduced above and using
(\ref{rho}) for the energy density, the WDW equation (\ref{H=0}) can
be written as 
\begin{equation}\label{wavefunct1}
\left\{\partial_u^2+\left(\frac{3}{2}\eta\right)^2\Omega_{d0}u^4\left[1+\sqrt{6}b\ln\left(u\right)\right]^2\right\}\Psi_1\left(u\right)=0, 
\end{equation}
where we have used the definitions given in (\ref{eta}). The
approximated WKB solution up to first order reads (see, for example,
the method used in \cite{Albarran:2015tga,Albarran:2015cda,Mathew}
and Appendix~A)
\begin{equation}\label{wavesol1}
\Psi_1(u)\approx\Omega_{d0}^{-\frac{1}{4}}\sqrt{\frac{2}{3\eta}}\frac{1}{u}\left[1+\sqrt{6}b\ln(u)\right]^{-\frac{1}{2}}\left\{D_1e^{i\frac{3\eta}{2}S_0(u)}+D_2e^{-i\frac{3\eta}{2}S_0(u)}\right\}, 
\end{equation}
where $D_1$ and $D_2$ are constants and 
\begin{equation}\label{S0}
S_0\left(u\right)=\sqrt{\Omega_{d0}}\int_{u_1}^{u}y^2\left[1+\sqrt{6}{ b}\ln\left(y\right)\right]dy=\frac{\sqrt{\Omega_{d0}}}{3}u^3\left\{1+\sqrt{6}b\left[\ln\left(u\right)-\frac{1}{3}\right]\right\}-\frac{\sqrt{\Omega_{d0}}}{3}u_1^3\left\{1+\sqrt{6}b\left[\ln\left(u_1\right)-\frac{1}{3}\right]\right\}. 
\end{equation}
In addition, $u_1$ is a large enough constant to ensure not only a
positive value of the above integral, but also to guarantee that the
system is well inside the DE domination regime, that is,
$\ln\left(u_1\right)\gg -1/(b\sqrt{6})$. Note that in the quantum treatment,
we disregard the contribution of DM by assuming a single component
through which the energy density is expressed. This is in
full agreement with the fact that by the time the classical abrupt event is
approached, DM contribution is negligible, see
Sec.~III.B.   
 From the inspection of (\ref{wavesol1}), we see that
the wave function vanishes for large values of $u$.
This is exactly the region where in the classical model the little rip
takes place. The DeWitt criterion is fulfilled, and the little rip is
avoided. It is interesting to note that this criterion is here
equivalent to the boundary condition that $\Psi_1\to 0$ for $a\to
\infty$ in analogy with the boundary condition usually imposed on the
Schr\"{o}dinger equation for bounded systems.

\subsubsection{Second quantization procedure (Laplace--Beltrami factor
  ordering): $\mathcal{\hat{H}}(a,\hat{\pi}_{a})\psi(a)=0$} 
This quantization procedure is based on the Laplace-Beltrami operator
which is the covariant generalization of the Laplacian operator in 
minisuperspace \cite{KieferQG}. The corresponding operator is
different depending on the involved degrees of freedom. For the case
of a single component described by a perfect fluid,  it is  written as
(cf.  for example Ref. \cite{Albarran:2015tga}) 
\begin{equation}\label{moment1}
\frac{\hat{\pi}_{a}^{2}}{a}=-\hbar^{2}\left[a^{-\frac{1}{2}}\frac{d}{da}\right]\left[a^{-\frac{1}{2}}\frac{d}{da}\right].
\end{equation}
To diagonalize the operator, we suggest the following change of variable
\begin{equation}\label{varchange}
z\equiv \left(\frac{a}{a_0}\right)^{\frac{3}{2}} \qquad , \qquad \frac{\hat{\pi}_{a}^{2}}{a}=-\frac{9}{4}\frac{\hbar^{2}}{{a}_0^3}\frac{d^{2}}{dz^{2}}.
\end{equation} 
Using this operator in the quantum version of
 (\ref{eq:wkbHamiltconstraint}) and multiplying by $\left[4\pi
   {a}_0^3/3G\hbar^2N\right]$,  we get the following dimensionless
 expression, 
\begin{equation}\label{Hamilq2}
\frac{4\pi {a}_0^3}{3G\hbar^2N}\hat{\mathcal {H}}=\partial_z^2 +\frac{8\pi^3 {a}_0^6}{3G\hbar^2}z^2\rho\left(z\right).
\end{equation}
Using (\ref{rho}) for the energy density,  the fundamental WDW
equation given in (\ref{H=0}) reduces to 
\begin{equation}\label{wavefunct2}
\left\{\partial_z^2+\eta^2\Omega_{d0}z^2\left[1+\sqrt{\frac{8}{3}}b \ln\left(z\right)\right]^2\right\}\Psi_2\left(z\right)=0,
\end{equation}
where the constants $\eta$ and $b$ are defined in (\ref{eta}).  The
approximated WKB  solution up to first order reads (see, for example,
\cite{Albarran:2015tga,Mathew}; for a summary, see also Appendix~A) 
\begin{equation}\label{wavesol2}
\Psi_2(a)\approx\Omega_{d0}^{-\frac{1}{4}}\sqrt{\frac{1}{\eta}}z^{-\frac{1}{2}}\left[1+\sqrt{\frac{8}{3}}b\ln(z)\right]^{-\frac{1}{2}}\left\{C_1e^{i\eta Q_0(z)}+C_2e^{-i\eta Q_0(z)}\right\},
\end{equation}
where $C_1$ and $C_2$ are constants and 
\begin{equation}\label{Q0}
Q_0\left(z\right)=\sqrt{\Omega_{d0}}\int_{z_1}^{z}y\left[1+\sqrt{\frac{8}{3}}
  b\ln\left(y\right)\right]dy=\frac{\sqrt{\Omega_{d0}}}{2}z^2\left\{1+\sqrt{\frac{8}{3}}b\left[\ln\left(z\right)-\frac{1}{2}\right]\right\}-\frac{\sqrt{\Omega_{d0}}}{2}z_1^2\left\{1+\sqrt{\frac{8}{3}}b\left[\ln\left(z_1\right)-\frac{1}{2}\right]\right\}. 
\end{equation}
Like in the first quantization procedure, we assume that $z_1$ is
large enough to ensure a positive value of the above integral; in
fact, it corresponds to the same scale factor $u_1$ that we used in the
previous quantization. 

As can be seen, the wave function vanishes for large values of $z$,
where the little rip takes place. Therefore, the DeWitt criterion is
again fulfilled; this can be seen as an indication that our results do
not depend on the chosen factor ordering.

Before concluding, we would like to highlight that both WKB solutions at first order can be related by
\begin{equation}
\frac{\left\vert\Psi_{1}\left(u\right)\right\vert^2}{\left\vert\Psi_{2}\left(z\right)\right\vert^2}= \frac{du}{dz}, \qquad \textrm{where}\qquad z=u^{\frac{3}{2}},
\end{equation}
where the equalities $D_1=C_1$ and $D_2=C_2$ have been
assumed.  At zero order, the two WKB solutions coincide as
$\left(3/2\right)S_0\left(u\right)=Q_0\left(z\right)$.


\subsection{Wheeler-DeWitt equation with a phantom scalar field}
For a system with a single (phantom) scalar field and a given potential, the
quantum Hamiltonian is written as
\cite{PMonizQC,KieferQG,Dabrowski:2006dd,Kamenshchik:2007zj,Albarran:2015cda} 
\begin{equation}\label{hamiltonian}
\mathcal{\hat{H}}=
N{a}_0^{-3}e^{-3x}\left\{\frac{\hbar^2}{4\pi^2}\left[\frac{\kappa^2}{6}\partial^2_x+\partial^2_{\phi}\right]+2\pi^2a_0^6e^{6x}V\left(\phi\right)\right\}. 
\end{equation}
Close to the little rip we can approximate the potential as
$V\left(\phi\right)\simeq b_1\phi^4$. Since
$\mathcal{\hat{H}}\Psi(x,\phi)=0$, we have 
\begin{equation}\label{Psi}
\left\{\frac{\hbar^2}{4\pi^2}\left[\frac{\kappa^2}{6}\partial^2_x+\partial^2_{\phi}\right]+\sigma e^{6x}\phi^4\right\}\Psi(x,\phi)=0,
\end{equation}
{\color{black}
where we have gathered all parameters in a single one called $\sigma$ which reads\footnote{{This
  Wheeler-DeWitt equation can be solved following the method
  introduced in Refs. [35-37] and, in particular, invoking the 
Symanzik scaling law. We briefly summarize this method in Appendix
C. We thank the referee for pointing out this method to us.}}
\begin{equation}
\sigma\equiv2\pi^2a_0^6b_1=\frac{27}{128}\pi^2a_0^6\kappa^2H_0^2\Omega_{*}.
\end{equation}
We next apply the following  change of variables:
\begin{eqnarray}\label{psirphi}
\begin{split}
\phi=r\left(z\right)\varphi, \ \ \ \  x=z,
\end{split}
\end{eqnarray}
where $r=r(z)$ is a function that  only depends on the new variable $z$. Consequently, we have
\begin{eqnarray}\label{psirphi2}
\begin{split}
\partial^2_{\phi}=&r^{-2}\partial^2_{\varphi}, \\ 
\partial^2_{x}=&\left(\frac{r^{\prime}}{r}\right)^2\left[\varphi^2\partial_{\varphi}^2+\varphi\partial_{\varphi}\right]-2\frac{{r^{\prime}}}{r}\varphi\partial_{\varphi}\partial_{z}+\left[\left(\frac{r^{\prime}}{r}\right)^2-\frac{r^{\prime\prime}}{r}\right]\varphi\partial_{\varphi}+\partial_{z}^2,
\end{split}
\end{eqnarray}
where primes  stands for derivatives with respect to $z$. Applying this change of  variable  and multiplying (\ref{Psi})  by
$r^2$, we  get 
\begin{equation}\label{Psi2}
\left\{\frac{\hbar^2\kappa^2}{24\pi^2}r^2\left[\left(\frac{r^{\prime}}{r}\right)^2\left[\varphi^2\partial_{\varphi}^2+\varphi\partial_{\varphi}\right]-2\frac{r^{\prime}}{r}\varphi\partial_{\varphi}\partial_{z}+\left[\left(\frac{r^{\prime}}{r}\right)^2-\frac{r^{\prime\prime}}{r}\right]\varphi\partial_{\varphi}+\partial_{z}^2\right]+\frac{\hbar^2}{4\pi^2}\partial^2_{\varphi}+\sigma e^{6z}r^6\varphi^4\right\}\Psi(z,\varphi)=0.
\end{equation}
Now, we choose $r\left(z\right)=e^{-z}$  with the aim to leave  the potential term  with a single dependence on the variable $\varphi$.
\begin{equation}\label{Psi3}
\left\{\frac{\hbar^2\kappa^2}{24\pi^2}e^{-2z}\left[\varphi^2\partial_{\varphi}^2+\varphi\partial_{\varphi}
+2\varphi\partial_{\varphi}\partial_{z}+\partial_{z}^2\right]+\frac{\hbar^2}{4\pi^2}\partial^2_{\varphi}+\sigma \varphi^4\right\}\Psi(z,\varphi)=0.
\end{equation}
We next  assume that in Eq.(\ref{Psi3}) some terms can be neglected under the presumption  
\begin{equation}\label{Psi3negl}
\frac{\hbar^2\kappa^2}{24\pi^2}e^{-2z}\left[\varphi^2\partial_{\varphi}^2+\varphi\partial_{\varphi}
+2\varphi\partial_{\varphi}\partial_{z}\right]\Psi(z,\varphi) \ \ \ll \ \ \frac{\hbar^2\kappa^2}{24\pi^2}e^{-2z}\partial_{z}^2\Psi(z,\varphi) \ \ ,\ \ \frac{\hbar^2}{4\pi^2}\partial^2_{\varphi}\Psi(z,\varphi)\ \ ,\ \ \sigma \varphi^4\Psi(z,\varphi),
\end{equation}
for large values of $z$ and $\varphi$ which is the regime where we want to solve the partial differential equation (\ref{Psi3}). This approximation must be justified after obtaining  the solutions for  $\Psi(z,\varphi)$ (see appendix \ref{justapp} for details). As can be seen, after disregarding these elements in Eq.(\ref{Psi3})  we have two terms whereby each
of them depends on a single variable.} Therefore, we can employ a
separation ansatz, and the wave function can be
written as a sum over products of two functions, 
{\color{black}
\begin{equation}\label{Psi4}
\Psi\left(z,\varphi\right)=\sum_k U_k\left(\varphi\right)C_k\left(z\right)q_k,
\end{equation}}
where $q_k$  denotes the  amplitude for each solution and  $k$  is a constant related to the ``energy'' of the system which characterizes the states described through the functions {\color{black}$C_k\left(z\right)$} and $U_k\left(\varphi\right)$. These functions, in turn, are  the solutions of the following differential equations
{\color{black}
\begin{equation}\label{Ck}
\left\{\frac{\hbar^2\kappa^2}{24\pi^2}\partial^2_z+ke^{2z}\right\}C_k(z)=0,
\end{equation}
\begin{equation}\label{Uk}
\left\{\frac{\hbar^2}{4\pi^2}\partial^2_{\varphi}+\sigma\varphi^4-k\right\}U_k(\varphi)=0.
\end{equation}
}
Equation (\ref{Uk}) corresponds to the inverted anharmonic oscillator in
quantum mechanics; see, for example, \cite{Bender06}.  
For {\color{black}$C_k(z)$,} we get exact solutions corresponding to  Bessel functions
with vanishing order: 
{\color{black}
\begin{itemize}
\item
For $k>0$
\begin{equation}\label{Cmsol1}
C_k(z)=C_{k1} J_{0}\left[\frac{2 \pi}{\hbar\kappa}\sqrt{6 k} \ e^z\right]+C_{k2} Y_{0}\left[\frac{2 \pi}{\hbar\kappa}\sqrt{6 k} \ e^z\right], 
\end{equation}
\item 
For $k<0$
\begin{equation}\label{Cmsol2}
C_k(z)=\tilde{C}_{k1} I_{0}\left[\frac{2 \pi}{\hbar\kappa}\sqrt{6 \vert k\vert} \ e^z\right]+\tilde{C}_{k2} K_{0}\left[\frac{2 \pi}{\hbar\kappa}\sqrt{6 \vert k\vert} \ e^z\right],
\end{equation}
\end{itemize}
where $C_{k1}$, $C_{k2}$ $\tilde{C}_{k1}$  and $\tilde{C}_{k2}$ are
constants. Since the functions $I_0(z)$ diverge for
$z\rightarrow\infty$ \cite{Abramow}, we choose $\tilde{C}_{k1}=0$ to
ensure that the wave function vanishes close to the little rip. 
For large values of $z$, we then get

\begin{itemize}
\item
For $k>0$
\begin{equation}\label{Cmlim1}
C_k(z)\sim \left[\frac{\hbar^2\kappa^2}{6\pi^4k}\right]^{\frac{1}{4}}e^{-\frac{z}{2}}\left\{C_{k1}\cos\left(\frac{2 \pi}{\hbar\kappa}\sqrt{6 \vert k\vert} \ e^z-\frac{\pi}{4}\right) +C_{k2}\sin\left(\frac{2 \pi}{\hbar\kappa}\sqrt{6 \vert k\vert} \ e^z-\frac{\pi}{4}\right) \right\},
\end{equation}
\item 
For $k<0$
\begin{equation}\label{Cmlim2}
C_k(z)\sim \tilde{C}_{k2}\left[\frac{\hbar^2\kappa^2}{96k}\right]^{\frac{1}{4}}e^{-\frac{z}{2}}.
\end{equation}
\end{itemize}
}
The second order differential equation for $U_k(\varphi)$ is 
more difficult to solve. Disregarding the  constant term $k$ in
(\ref{Uk}), which is equivalent to finding the solution for $k=0$, it
can be written as (see the appendix A)\footnote{Naively, we expect $k$
to be irrelevant close to the little rip where $\varphi$ gets very
large values; see below for a rigorous justification of this observation.} 
  
{\color{black}
\begin{equation}\label{U0sol}
U(\varphi)=\sqrt{\varphi}\left\{U_{1} J_{\frac{1}{6}}\left[\frac{2\pi\sqrt{\sigma}}{3\hbar}\varphi^3\right]+U_{2} J_{-\frac{1}{6}}\left[\frac{2\pi\sqrt{\sigma}}{3\hbar}\varphi^3\right]\right\}, 
\end{equation}
where $U_{1}$ and $U_{2}$ are integration constants. For large values of $\varphi$, we have
\begin{equation}\label{U0sollim}
U(\varphi)\sim \sqrt{\frac{6\hbar}{2\pi^2\sigma^{\frac{1}{2}}}}\frac{1}{\varphi}\left\{U_{1} \cos\left(\frac{2\pi\sqrt{\sigma}}{3\hbar}\varphi^3-\frac{\pi}{3}\right) +U_{2} \sin\left(\frac{2\pi\sqrt{\sigma}}{3\hbar}\varphi^3-\frac{\pi}{3}\right) \right\},
\end{equation}
and the wave function vanishes asymptotically. It is worth notice that for small values of the argument  $\left(2\pi\sqrt{\sigma}/3\hbar\right)\varphi^3$ in Eq.(\ref{U0sollim}) we have
\begin{equation}\label{Usmall}
U(\varphi)\sim U_1\left(\frac{\pi\sqrt{\sigma}}{3\hbar}\right)^{\frac{1}{6}}\frac{\varphi}{\Gamma\left(\frac{7}{6}\right)}-U_2\left(\frac{\pi\sqrt{\sigma}}{3\hbar}\right)^{-\frac{1}{6}}\frac{\Gamma\left(\frac{1}{6}\right)}{\pi} ~ {\color{red}.}
\end{equation}
This limit seems to correspond to a regime where $\sigma$ (which is proportional to the parameter $\Omega_{*}$, i.e. quadratic in $A^2$) is small enough to ensure  infinitesimal  values of the argument  in Eq.(\ref{U0sol}) even for  large values of $\varphi$. It turns out that the term proportional to $U_2$ in Eq.(\ref{Usmall}) is not well defined when $\sigma$ or $A$ vanishes. This might indicate that the wave function for the $\Lambda$CDM universe is not well defined. However, this is not the case because when $\sigma$ or $A$ approaches  zero $V\left(\phi\right)$ should be the one given in Eq.(\ref{Vasmall}) rather than we used and defined in Eq.(\ref{potphi}). In addition, the solution (\ref{Usmall}) was obtained after disregarding the term $k$ in Eq.(\ref{Uk}) which cannot be  ignored in the case of small $\sigma\varphi^4$. }

After performing the approximation ($k\ll[4\pi^2\sigma/\hbar^2]\varphi^4$) in (\ref{Uk}), we
can find an exact solution, but in return, we lose the information of
$k$ in $U_k(\varphi)$. A  simple way to obtain an approximated wave
function keeping the contribution of $k$ is via  the WKB
approximation, the expression for the approximated wave function up to
first order is given by 
{\color{black}
\begin{equation}\label{UmWKBsol}
U(\varphi)\simeq\left[\frac{4\pi^2}{\hbar^2}\left(\sigma\varphi^4-k\right)\right]^{-\frac{1}{4}}\left\{U_{k1} e^{iS_0\left(\varphi\right)}+U_{k2} e^{-iS_0\left(\varphi\right)}\right\}, 
\end{equation}
where $U_{k1}$ and $U_{k2}$ are constants and}
\begin{equation}\label{Soyjust}
S_0(\varphi)=\frac{2 \pi}{\hbar}\int_{\varphi_1}^{\varphi}\sqrt{\sigma y^4-k} \ dy ~,
\end{equation}
where $\varphi_1$ is large enough to ensure a purely real solution
even for positive values of $k$
($0<\sigma\varphi_1^4-k$). The latter integral can be
expressed as follows (see pages 128 and 129 of \cite{INTandSER})  
{\color{black}
\begin{itemize}
\item for $0<k$
\begin{equation}\label{intFkpos}
\int_{\varphi_1}^{\varphi}\sqrt{\sigma y^4-k} \
dy=\frac{2 \pi}{3\hbar}\left.\left\{y\left(\sigma y^4- k\right)^{\frac{1}{2}}-\sqrt{2}\frac{ k^{\frac{3}{2}}}{\sigma^{\frac{1}{4}}}
    {F}\left[\arccos\left(\frac{k^{\frac{1}{4}}}{\sigma^{\frac{1}{4}}y}\right),\frac{1}{\sqrt{2}}\right]\right\}\right\vert^{\varphi}_{\varphi_1}, 
\end{equation} 
\item for $k<0$
\begin{equation}\label{intFkneg}
\int_{\varphi_1}^{\varphi}\sqrt{\sigma y^4-k} \
dy=\frac{2 \pi}{3\hbar}\left.\left\{y\left(\sigma y^4+\left\vert k\right\vert\right)^{\frac{1}{2}} +\frac{\left\vert k\right\vert^{\frac{3}{2}}}{\sigma^{\frac{1}{4}}}{F}\left[\arccos\left(\frac{\sqrt{\left\vert k
            \right\vert}-\sqrt{\sigma}y^2}{\sqrt{\left\vert k
            \right\vert}+\sqrt{\sigma}y^2}\right),
      \frac{1}{\sqrt{2}}\right]\right\}\right\vert^{\varphi}_{\varphi_1}, 
\end{equation}

\end{itemize} 
}
\noindent where the function ${F}\left[h(y),d\right]$ is an elliptic
integral of the first kind with argument $h(y)$ and elliptic modulus
$d$. Note that for $k=0$ we recover the asymptotic solution given by
the Bessel functions (\ref{U0sol}). For large values of $\varphi$ the
performed WKB approximation and the found Bessel functions has the
same asymptotic  behavior,  in this limit, no matter what is the value
of $k$. Therefore, for very large values of $\varphi$ we can write 
\begin{equation}\label{Psilimit}
\Psi\left(z,\varphi\right)\simeq U\left(\varphi\right)\sum_kC_k\left(z\right)q_k,
\end{equation}
In any case, the resulting wave function has two oscillatory terms modulated by a function which  goes to zero for large values.  Returning to the initial variables, for $z\rightarrow\infty$ and $\varphi\rightarrow\infty$ limits the wave function decrease as
\begin{equation}\label{Psilimit2}
\Psi(x,\phi)\sim\left[\phi e^{\frac{3}{2}x}\right]^{-1}.
\end{equation}
Therefore, the wave function vanishes close to the little rip, fulfilling the DeWitt boundary condition. 

 
\section{Conclusions}
\label{Sec5}

A central issue in any theory of quantum gravity is the avoidance of
classical singularities. At the present state of the field, this
cannot be done in any sense close to the rigour of the classical
singularity theorems. The hope is thus to get some insight from 
{\color{black} suitable } models for which concrete results 
can be obtained. As a heuristic sufficient 
(though not necessary) criterion of singularity avoidance,
one can employ the DeWitt criterion of vanishing wave function.
The applicability of this criterion has already been studied for a wide
class of classical singularities. In the present paper, we have
completed the discussion by studying the situation of the little rip,
which is strictly speaking not a singularity, like it is the case of a big rip, but an abrupt event,
though it shares some features with it. We have studied the
two situations of a perfect fluid and of a phantom scalar field,
the first being a phenomenological, the second a more fundamental
dynamical model.  
A phantom field (field with negative kinetic energy) is needed in
order to implement the equation of state leading to a little rip. 
We have found that the DeWitt criterion can indeed be applied in both
cases and that the little rip can thus be avoided. We should emphasize
that models such as these, although looking purely academic at first
glance, are supported by data \cite{Jimenez:2016sgs}. 
If indeed true, the future of the Universe would end in a full
quantum era (without classical observers), in full analogy to its
quantum beginning.

{\color{black}

Some words about the applicability of the DeWitt criterion
(\cite{DeWitt}, Eq. (6.31)) are in order. It is based on the heuristic
extrapolation of the quantum mechanical probability interpretation
(based on the Schr\"odinger inner product) to
quantum cosmology. But since the Wheeler-DeWitt equation is of
hyperbolic nature (with and without matter),
and thus resembles a Klein-Gordon equation, one might think that
a Klein-Gordon inner product would be more appropriate. This is, however,
not the case, because it was proven that one cannot separate positive and
negative frequencies in the Wheeler-DeWitt equation, and thus one
is faced with the problem of negative probabilites; see, for example,
\cite{KieferQG}, Sec. 5.2.2 for a discussion and references.
This problem can perhaps
be avoided by going to ``third quantization'', but this is a framework different
from the present one.  Our point of view here is that an inner product
of the Schr\"odinger type can be used in quantum
cosmology, even if the situation in the full theory is
unclear\footnote{The formalism of full loop quantum gravity, for example,
employs a Schr\"odinger inner product.} and
even if this poses the danger of not allowing normalizable
solutions. At least in the models hitherto considered,
this inner product can be implemented and the DeWitt criterion
can be applied.}

In our paper, we have restricted ourselves to the
minisuperspace approximation. The real Universe is, however, not
homogeneous, so one possible extension of our work is the inclusion
of (scalar and tensor) perturbations and solving the WDW equation near and at the region 
of the little rip. The full quantum state then describes an
entanglement between the minisuperspace part and the perturbations. Tracing
out the perturbation part from the full state leads to a density
matrix $\rho$ for the minisuperspace part. If the interaction with the
perturbations leads to a suppression of the off-diagonal elements in
$\rho$, one can interpret this as an effective quantum-to-classical
transition or decoherence for the background. Decoherence in quantum
cosmology was discussed in detail for many situations; see, for
example, \cite{BKCM99} and the references therein. One might expect
that close to the little rip region, decoherence stops and
quantum interferences become important, enabling the DeWitt criterion
to be fulfilled there, as discussed in our paper. Genuine quantum
effects have also shown to be important near the turning point of a
classically recollapsing universe \cite{KZ95}. 
 We hope to address these and other issues in future publications.


\section{Acknowledgments}

The work of IA is supported by a Santander-Totta fellowship ``Bolsas
de Investiga\c{c}\~{a}o Faculdade de Ci\^{e}ncias (UBI)-Santander
Totta''. The work of MBL is supported by the Portuguese Agency
“Funda\c{c}\~ao para a Ci\^encia e Tecnologia” through an Investigador
FCT Research contract, with reference IF/01442/2013/
CP1196/CT0001. She also wishes to acknowledge the partial support from
the Basque government Grant No. IT592-13 (Spain) and  FONDOS FEDER
under grant  FIS2014-57956-P (Spanish government). This research work
is supported by the Portuguese grant UID/MAT/00212/2013.


\appendix

\section{WKB approximation}
\label{appendix}


We next review briefly the WKB method for second order differential equation 
\begin{equation}\label{eq:wdwrWKBia}
\left[\frac{d^{2}}{dy^{2}}+V_{\rm eff}\left(y\right)\right]\psi(y)=0 ~,
\end{equation}
where $y$ is defined such that it is  a dimensionless degree of
freedom and where the effective potential can be written as  
\begin{equation}\label{veffy}
V_{\rm eff}\left(y\right)=\tilde{\eta}^2 g(y) ~.
\end{equation}
Moreover, $\tilde{\eta}$ is a dimensionless parameter related with the constants of the system. The general expression for the WKB approximated solution (up to first order) reads \cite{Mathew}
\begin{equation}\label{wkbfirstorder}
\psi(y)\approx \left[-\tilde{\eta}^2g\left(y\right)\right]^{-\frac{1}{4}}\left[B_{1}e^{iS_0(y)}+B_{2}e^{-iS_0(y)}\right],
\end{equation}
where $B_1$ and $B_2$ are constants and
\begin{equation}\label{Soy}
S_0(y)=\tilde{\eta}\int_{y_1}^{y}{\sqrt{g(y)}dy}.
\end{equation}
The solution (\ref{wkbfirstorder}) is valid as long as the inequality
\begin{equation}\label{wkbjust<1}
\frac{1}{\tilde{\eta}^{2}}\left|\frac{5\dot{g}^{2}(y)-4\ddot{g}(y)g(y)}{16g^{3}(y)}\right|\ll1,
\end{equation}
is fulfilled. When the left hand side of Eq. (\ref{wkbjust<1}) goes to zero we can be sure that the  behavior of the exact solution in this regime matches almost perfectly with the WKB approximation.
The approximations used in  section \ref{WDWsection} for the first and the second quantization procedure, corresponds to an effective potential whose general shape reads
\begin{equation}\label{g(y)}
V_{\rm eff}\left(y\right)\equiv\tilde{\eta}\ y^{n}\left[1+\gamma\ln\left(y\right)\right],
\end{equation}
where for each quantization procedure we have that:
\begin{itemize}
\item
First quantization procedure:
\begin{equation}\label{etatilde,n,gamma1}
\tilde{\eta}=\frac{2}{3}\sqrt{\Omega_{d0}}\eta  \qquad , \qquad  n=2  \qquad , \qquad  \gamma=\sqrt{6}b.
\end{equation}
\item
Second quantization procedure:
\begin{equation}\label{etatilde,n,gamma2}
\tilde{\eta}=\sqrt{\Omega_{d0}}\eta\qquad , \qquad  n=1  \qquad , \qquad  \gamma=\sqrt{\frac{8}{3}}b.
\end{equation}
\end{itemize}
The necessary condition for the approximation to be valid reads
\begin{equation}\label{wkbjustpot}
\frac{1}{\tilde{\eta}^2}\left\vert\frac{1}{16 y^{n+2}\left[1+\gamma\ln\left(y\right)\right]^2}\left\{n^2\left(9-4n\right)+\frac{4\gamma n\left(1-6n-2\gamma\right)}{\left[1+\gamma\ln\left(y\right)\right]}+\frac{\gamma^2 \left(36-32n-16\gamma\right)}{\left[1+\gamma\ln\left(y\right)\right]^2}\right\}\right\vert\ll1.
\end{equation}
For both of the cases  mentioned above, the little rip occurs for large values of the variable $y$, where the latter expression goes to zero when $y\rightarrow\infty$; i.e. the WKB approximated solution is valid.

{\color{black}
\section{Justification for the approximation done in Eq. (\ref{Psi3})}\label{justapp}

The approximation done for the differential equation (\ref{Psi3}) consists into disregarding the first three terms after the change of variable realized in (\ref{psirphi}). Once these terms are neglected, the resulting differential equation is separable and the approximation is valid if 
\begin{equation}\label{Psi3negljust}
\frac{\hbar^2\kappa^2}{24\pi^2}e^{-2z}\left[\varphi^2\partial_{\varphi}^2+\varphi\partial_{\varphi}
+2\varphi\partial_{\varphi}\partial_{z}\right]C\left(z\right)U\left(\varphi\right) \ll \frac{\hbar^2\kappa^2}{24\pi^2}e^{-2z}\partial_{z}^2C\left(z\right)U\left(\varphi\right)  , \frac{\hbar^2}{4\pi^2}\partial^2_{\varphi}C\left(z\right)U\left(\varphi\right),\sigma \varphi^4C\left(z\right)U\left(\varphi\right).
\end{equation}
As a result of the realized approximation, the last two terms in the rhs of the above  inequality have the same order of magnitude for large values of $z$ and $\varphi$ or $\phi$ and $x$ (See Eq.(\ref{accept})).
In fact, the dominant terms that we keep reads 
\begin{eqnarray}\label{accept}
\begin{split}
&\frac{\hbar^2}{4\pi^2}\partial^2_{\varphi}C\left(z\right)U\left(\varphi\right),\sigma \varphi^4C\left(z\right)U\left(\varphi\right)\sim\frac{\hbar}{2\pi}\left(\frac{\kappa^2}{6k}\right)^{\frac{1}{4}}\sigma^{\frac{3}{4}}\varphi^{3}e^{-\frac{1}{2}z}=\frac{\hbar}{2\pi}\left(\frac{\kappa^2}{6k}\right)^{\frac{1}{4}}\sigma^{\frac{3}{4}}\phi^{3}e^{\frac{5}{2}x},\\
&\frac{\hbar^2\kappa^2}{24\pi^2}e^{-2z}\partial_{z}^2C\left(z\right)U\left(\varphi\right)\sim\frac{\hbar}{2\pi}\left(\frac{\kappa^2k^3}{6}\right)^{\frac{1}{4}}\sigma^{-\frac{1}{4}}\varphi^{-1}e^{-\frac{1}{2}z}=\frac{\hbar}{2\pi}\left(\frac{\kappa^2k^3}{6}\right)^{\frac{1}{4}}\sigma^{-\frac{1}{4}}\phi^{-1}e^{-\frac{3}{2}x},
\end{split}
\end{eqnarray}
while the  neglected terms evolve asymptotically as 
\begin{eqnarray}\label{negec}
\begin{split}
&\frac{\hbar^2\kappa^2}{24\pi^2}e^{-2z}\varphi^2\partial_{\varphi}^2C\left(z\right)U\left(\varphi\right)\sim \frac{\hbar}{12\pi}\left(\frac{\kappa^{10}}{6k}\right)^{\frac{1}{4}}\sigma^{\frac{3}{4}}\varphi^{5}e^{-\frac{5}{2}z}=\frac{\hbar}{12\pi}\left(\frac{\kappa^{10}}{6k}\right)^{\frac{1}{4}}\sigma^{\frac{3}{4}}\phi^{5}e^{\frac{5}{2}x},\\
&\frac{\hbar^2\kappa^2}{24\pi^2}e^{-2z}\varphi\partial_{\varphi}
C\left(z\right)U\left(\varphi\right)\sim\frac{\hbar^2}{24\pi^2}\left(\frac{\kappa^{10}}{6k}\right)^{\frac{1}{4}}\sigma^{\frac{1}{4}}\varphi^{2}e^{-\frac{5}{2}z}=\frac{\hbar^2}{24\pi^2}\left(\frac{\kappa^{10}}{6k}\right)^{\frac{1}{4}}\sigma^{\frac{1}{4}}\phi^{2}e^{-\frac{1}{2}x},\\
&\frac{\hbar^2\kappa^2}{12\pi^2}e^{-2z}\varphi\partial_{\varphi}\partial_{z}C\left(z\right)U\left(\varphi\right)\sim \frac{\hbar}{6\pi}\left(\frac{3\kappa^{6}k}{8}\right)^{\frac{1}{4}}\sigma^{\frac{1}{4}}\varphi^{2}e^{-\frac{3}{2}z}= \frac{\hbar}{6\pi}\left(\frac{3\kappa^{6}k}{8}\right)^{\frac{1}{4}}\sigma^{\frac{1}{4}}\phi^{2}e^{\frac{1}{2}x}.
\end{split}
\end{eqnarray}
Therefore,   in order to obtain the compliance region of the realized  approximation we compare the smallest of the saved terms with the largest between the neglected ones, that is
\begin{eqnarray}
\begin{split}
\frac{\hbar}{12\pi}\left(\frac{\kappa^{10}}{6k}\right)^{\frac{1}{4}}\sigma^{\frac{3}{4}}\phi^{5}e^{\frac{5}{2}x}&\ll
\frac{\hbar}{2\pi}\left(\frac{\kappa^2k^3}{6}\right)^{\frac{1}{4}}\sigma^{-\frac{1}{4}}\phi^{-1}e^{-\frac{3}{2}x}, \\
\Longrightarrow \quad  \frac{\kappa^2\sigma}{6k}\phi^{6}e^{4x}&\ll 1.
\end{split}
\end{eqnarray}
Finally, the realized approximation is valid as long as  $\left(\kappa^2\sigma/6k\right)\phi^{6}e^{4x}\ll 1$. This means that for sufficiently small values of $\sigma$; i.e. 
for small values of\footnote{Small values of $A$ implies small deviations of our model from the $\Lambda$CDM scenario.} $A$, which is indeed the observationally preferred situation, 
and large values of $x$; i.e. $x\gg 1$ (but finite), the approximation we have used is valid. 

}

\section{Scalar field eigenstates and Symanzik scaling behavior}
\label{appendix2}

{ In this Appendix we analyze Eq. (\ref{Psi}) in the context of the Symanzik scaling law, following the results found in \cite{Symanzik1,Symanzik2,Symanzik3}.
We start by performing, in the aforementioned equation, the following change of variables, 
\begin{equation}
x=c_{1}\bar{x} \;,\quad\phi=c_{2}\bar{\phi}\;,
\end{equation}
where $c_1$ and $c_2$ are constants. We obtain
\begin{equation}
\left\{ \frac{\hbar^{2}}{4\pi^{2}}\left[\frac{\kappa^{2}}{6c_{1}^{2}}\partial_{\bar{x}}^{2}+\dfrac{1}{c_{2}^{2}}\partial_{\bar{\phi}}^{2}\right]+\sigma e^{6c_{1}\bar{x}}\, c_{2}^{4}\bar{\phi}^{4}\right\} \Psi(\bar{x},\,\bar{\phi})=0 \; ,
\label{C2}
\end{equation}
where by imposing
\begin{equation}
c_{1}  =\dfrac{\hbar}{2\sqrt{6}\pi}\kappa \; , \quad
c_{2}  =i\dfrac{\hbar}{2\pi} \; ,
\end{equation}
we get
\begin{equation}
\left[\partial_{\bar{x}}^{2}-\partial_{\bar{\phi}}^{2}+\dfrac{\hbar^{4}}{16\pi^{4}}\sigma e^{\frac{\sqrt{6}}{2\pi}\hbar\kappa\bar{x}}\bar{\phi}^{4}\right]\Psi(\bar{x},\,\bar{\phi})=0 \; ,
\label{C4}
\end{equation}
which is precisely given by Eq. (1) of Ref. \cite{Symanzik1}. As done in \cite{Symanzik1}, we conclude that the general solutions of (\ref{C4}) can be expressed as
\begin{equation}
\Psi(\bar{x},\,\bar{\phi})=\sum_{n=0}^{+\infty}A_{n}\left(\bar{x}\right)\Phi_{n}\left(\bar{x},\,\bar{\phi}\right) \; ,
\end{equation}
which can be rewritten in a vectorial notation as
\begin{equation}
\Psi(\bar{x},\,\bar{\phi})=\mathbf{\Phi}^{T}\left(\bar{x},\,\bar{\phi}\right)\cdot\mathbf{A}\left(\bar{x}\right) \; ,
\end{equation}
where the scalar part wave functions satisfy
\begin{equation}
\left[\partial_{\bar{\phi}}^{2}+E_{n}\left(\bar{x}\right)-\dfrac{\hbar^{4}}{16\pi^{4}}\sigma e^{\frac{\sqrt{6}}{2\pi}\hbar\kappa\bar{x}}\bar{\phi}^{4}\right]\Phi_{n}\left(\bar{x},\,\bar{\phi}\right)=0 \; .
\end{equation}
Notice that the scalar part solutions depend on the scale factor. Using the Symanzik scaling law \cite{Symanzik1,Symanzik2,Symanzik3}, we have that
\begin{equation}
\begin{aligned}\Phi_{n}\left(\bar{x},\,\bar{\phi}\right) & =\left[\dfrac{\hbar^{4}}{16\pi^{4}}\sigma e^{\frac{\sqrt{6}}{2\pi}\hbar\kappa\bar{x}}\right]^{\frac{1}{12}}f_{n}\left(\bar{\chi}\right)\,,\\
E_{n}\left(\bar{x}\right) & =\left[\dfrac{\hbar^{4}}{16\pi^{4}}\sigma e^{\frac{\sqrt{6}}{2\pi}\hbar\kappa\bar{x}}\right]^{\frac{1}{3}}\varepsilon_{n}\,,
\end{aligned}
\end{equation}
where
\begin{equation}
\bar{\chi}=\left[\dfrac{\hbar^{4}}{16\pi^{4}}\sigma e^{\frac{\sqrt{6}}{2\pi}\hbar\kappa\bar{x}}\right]^{\frac{1}{12}}\bar{\phi} \; .
\end{equation}
Furthermore, the vectorial scalar field wave equation can be used to define a coupling matrix $\mathbf{\Omega}$ as \cite{Symanzik1,Symanzik2,Symanzik3}
\begin{equation}
\dfrac{\partial \mathbf{\Phi}}{\partial\bar{x}}=\mathbf{\Omega}\mathbf{\Phi}\left(\bar{x},\,\bar{\phi}\right)\,.
\end{equation}
Given that $\left\lbrace\Phi_{n}\right\rbrace$ are an orthonormal basis, we can conclude that
\begin{equation}
\Omega_{mn}=\dfrac{\varepsilon_{m}-\varepsilon_{n}}{4}\int d\bar{\chi}\,\bar{\chi}^{2}f_{m}\left(\bar{\chi}\right)f_{n}\left(\bar{\chi}\right) \; .
\end{equation}

}

\end{document}